\documentclass[aps,prb,reprint,showpacs,floatfix]{revtex4-1}

\usepackage{color}
\usepackage{graphicx}

\begin{document}


\title{Blind deconvolution of density-matrix renormalization-group spectra}


\author{M. Paech}
\author{E. Jeckelmann}

\affiliation{Institut f\"ur Theoretische Physik, Leibniz Universit\"at Hannover, Appelstrasse 2, D-30167 Hannover, Germany}

\date{\today}

\begin{abstract}
We present a numerical method for calculating piecewise smooth spectral 
functions of correlated quantum systems in the thermodynamic limit
from the spectra of finite systems computed using the dynamical or
correction-vector density-matrix renormalization group method.
The key idea is to consider this problem as a blind deconvolution
with an unknown kernel which causes both a broadening and 
finite-size corrections of the spectrum.
In practice, the method reduces to a least-square optimization
under non-linear constraints which enforce the positivity and
piecewise smoothness of spectral functions.
The method is demonstrated on the single-particle density of states
of one-dimensional paramagnetic Mott insulators represented by the half-filled Hubbard model
on an open chain. Our results confirm that the density of states has a step-like shape but no square-root singularity at the
spectrum onset.
\end{abstract}


\maketitle



%



\section{Introduction}

The dynamical density--matrix renormalization group (DMRG)~\cite{jec00,jec02}
and the
closely related correction--vector DMRG~\cite{kue99,pat99}
have been widely used in the last decade to compute the dynamical correlation 
functions and
spectral functions
of low-dimensional strongly correlated quantum systems.~\cite{jec08a,jec08b}
Although more powerful DMRG approaches have been developed 
recently,~\cite{whi08,wei09,hol11,dar11}
dynamical DMRG (DDMRG) often  remains the method of choice because it offers 
two 
practical advantages over the other approaches:
it is simpler and it can be easily parallelized.
For instance, it has been recently shown that DDMRG allows us to investigate
features with small spectral weights such as power-law pseudo-gaps in 
Luttinger liquids.~\cite{jec13}
The main drawback of DDMRG is that it always yields the convolution of the 
desired spectrum with a Lorentzian distribution of finite
width.  Therefore, the true spectrum can only be obtained through a 
deconvolution of the DDMRG spectrum.
(In principle, there are some methods to get around this 
problem~\cite{kue99,pet11} but they are rarely used in practice.)

Deconvolution is a typical ill-conditioned inverse problem, 
however.~\cite{nr,ole09}
A direct solution of the deconvolution equation usually yields a very noisy
and thus useless spectrum.
Nevertheless, various regularization methods have been successfully used to
deconvolve 
DDMRG spectra for one-dimensional systems and quantum impurity 
problems.~\cite{wei09,geb03,nis04a,nis04b,raa04,raa05,jec07,sch10,ulb10}
Astonishingly, some of these deconvolution methods even allow us to bypass the 
finite-size scaling analysis and 
to obtain the piecewise smooth spectrum of an infinite systems directly
from a broadened finite-system DDMRG spectrum.
Unfortunately, regularization also smooths out the sharp features of the true 
spectrum. 
This is a serious issue as the spectra of 
one-dimensional systems and quantum impurities often exhibit very
interesting (power-law) singularities.

In this paper we present a method, which allows us to 
determine sharp spectral features in the thermodynamic limit
starting from a broadened finite-system DDMRG spectrum. 
For this purpose we consider the
extrapolation to the thermodynamic limit and the deconvolution for the 
Lorentzian kernel
to be a single blind deconvolution,~\cite{ole09,cam07} i.\,e.\ an inverse 
problem with an unknown kernel including
both the Lorentzian broadening and the finite-size effects. 
The key idea to preserve sharp spectral features in a piecewise smooth spectrum
is
to impose a minimal distance $d$ between extrema of the deconvolved spectrum.
To illustrate our method we investigate the single-particle density of states 
(DOS) of one-dimensional paramagnetic
Mott insulators represented by the half-filled Hubbard model.~\cite{hub63}
We confirm that this DOS has the step-like onset predicted by field-theoretical 
studies~\cite{ess02} at least at weak to intermediate coupling up to $U=8t$.

\section{Model and observable}

The Hubbard model~\cite{hub63}
with on-site interaction $U\geq 0$ and nearest-neighbor hopping $t$
is a basic lattice model for the physics
of strongly interacting electrons, in particular the Mott metal-insulator
transition.~\cite{mot90,geb97}
At half filling (i.\,e., the number of electrons equals the number of sites $N$)
the ground state
is a Mott insulator for strong interaction $U/t$, while it is a Fermi
gas in the non-interacting limit $U=0$. 
The Hamiltonian of the Hubbard model is defined by  
\begin{eqnarray}
H & = &- t \sum_{\langle ij\rangle, \sigma}
\left (  c^{\dagger}_{i,\sigma} c^{\phantom{\dagger}}_{j,\sigma} 
+ c^{\dagger}_{j,\sigma} 
c^{\phantom{\dagger}}_{i,\sigma}  \right ) 
- \mu \sum_i n_{i}  \nonumber \\
&& + U \sum_i n_{i,\uparrow} n_{i,\downarrow} 
\label{eq:hamiltonian}
\end{eqnarray}
where the operator $c^{\dagger}_{i,\sigma}$ 
($c^{\phantom{\dagger}}_{i,\sigma}$)
creates (annihilates) an electron with spin $\sigma$ on the site $i$,
$n_{i,\sigma} = c^{\dagger}_{i,\sigma} c^{\phantom{\dagger}}_{i,\sigma}$,
and $n_i = n_{i,\uparrow} + n_{i,\downarrow}$.
The first sum runs over all pairs $\langle ij\rangle$
of nearest-neighbor sites while the other
two sums run over all sites $j$.
Here we will only consider half-filled systems and
thus set the chemical potential $\mu = \frac{U}{2}$ 
to have electron-hole-symmetric spectra and a Fermi energy 
$\epsilon_{\rm F} = 0$.

The bulk single-particle DOS $D(\epsilon)$ can be measured experimentally 
using photoemission spectroscopy
or scanning tunneling spectroscopy. 
Theoretically, it can be defined as the average of the
local DOS 
\begin{equation}
D(\epsilon) = \frac{1}{2N} \sum_{j,\sigma} D_{\sigma}(j,\epsilon)
\end{equation}
where the sum runs over both spins and all sites $j$ in the lattice,
while
$D_{\sigma}(j,\epsilon)$ is the local single-particle DOS at site $j$ for 
spin $\sigma$ and  can be 
calculated using 
\begin{equation}
D_{\sigma}(j,\epsilon) = \sum_{n} \vert \langle n \vert
c^{\dagger}_{j,\sigma} 
\vert 0 \rangle \vert^2 \ \delta(\epsilon - E_{n} +E_0)
\end{equation}
for $\epsilon \geq 0$
and \begin{equation}
D_{\sigma}(j,\epsilon) = \sum_{n} \vert \langle n \vert c_{j,\sigma} 
\vert 0 \rangle \vert^2 \ \delta(\epsilon + E_{n} -E_0)
\end{equation}
for $\epsilon \leq 0$.
Here $\vert n \rangle$ denotes the eigenstates of the Hamiltonian $H$ and 
$E_n$ their eigenenergies in the Fock space. 
The ground state for the chosen number of particles corresponds to $n=0$.
The total spectral weight is 
\begin{equation}
\int_{-\infty}^{+\infty} d \epsilon \ D_{\sigma}(j,\epsilon) = 1 .
\label{eq:sum}
\end{equation}
We will consider only lattice geometry for which the 
Hamiltonian~(\ref{eq:hamiltonian}) is invariant under the electron-hole
transformation
$c^{\dag}_{j,\sigma} \leftrightarrow  \pm c^{\phantom{\dag}}_{j,\sigma}$. 
Therefore, for half filling the density of states is symmetric, 
$D_{\sigma}(j,\epsilon) = D_{\sigma}(j,-\epsilon)$.
If the system is translation invariant, the bulk DOS and the local DOS are 
identical. For DMRG simulations, however, open boundary
conditions are preferred to periodic boundary conditions. 
In this case, the bulk DOS can be identified with the local
DOS on one of the two equivalent middle sites of the system, i.\,e.\ as far as possible from the system 
boundaries.~\cite{jec13}

\section{Deconvolution}

Inverse problems such as (blind) deconvolutions~\cite{nr,ole09} occur
in many scientific fields and are among the most challenging numerical
computations.  
Experimental measurements and computer simulations often yield approximations
of the true quantities which are measured or computed, respectively.
It is often assumed that the deviations from exact results 
can be modelled by a convolution with
a smoothing function and an additive noise
due to the finite accuracy and resolution of the measurement or simulation
process. 
A typical example of a blind deconvolution 
is the reconstruction of an original signal from
a degraded copy using incomplete information about the
degradation process.~\cite{ole09,cam07}
Here we want to compute sharp spectral features in the piecewise smooth 
spectrum of an infinite system from
a broadened finite-system spectrum calculated with DDMRG. In this section we 
first show that this task
can be formulated as a blind deconvolution problem, then present an algorithm 
for solving it.  

\subsection{Blind deconvolution problem}

Let $S^{(N)}(\epsilon)$ be a spectrum of a finite lattice model with $N$ 
sites. 
This spectrum is a Dirac-comb (a finite sum of Dirac-peaks)
\begin{equation}
S^{(N)}(\epsilon)  = \sum_{n} S^{(N)}_{n} \ \delta \left ( 
\epsilon - E^{(N)}_{n} \right ) 
\end{equation}
where the sum runs over all Hamiltonian eigenstates $\vert n \rangle$ 
which contributes to the spectrum, i.\,e.\ with a nonzero spectral 
weight $S^{(N)}_{n}$. 
Here $E^{(N)}_n$ denotes the corresponding excitation energies.
This spectrum can be broadened with a Lorentzian distribution of width $\eta$
\begin{equation}
L_{\eta}(\epsilon) = \frac{1}{\pi} \frac{\eta}{\epsilon^2+ \eta^2}
\label{eq:lorentz-kernel}
\end{equation}
to obtain a smooth spectral function
\begin{eqnarray}
S^{(N)}_{\eta}(\epsilon) & = &
\int d\omega  \ S^{(N)}(\omega) \  L_{\eta}(\epsilon-\omega)  \nonumber \\ 
& = & \frac{1}{\pi} \sum_{n} S^{(N)}_{n} \frac{\eta}{ \left (
\epsilon-E^{(N)}_n \right )^2 
+ \eta^2}  . 
\label{eq:spec-finite}
\end{eqnarray}
With the DDMRG method we can calculate this spectrum for
a discrete set of excitation energies
$\{\epsilon_{\alpha}; \alpha=1,\dots,M \}$.
As numerical calculations are always affected by errors,
DDMRG actually yields values $W^{(N)}_{\eta}(\epsilon_{\alpha})$  
which are related to the true spectral function by 
\begin{equation}
W^{(N)}_{\eta}(\epsilon_{\alpha}) = S^{(N)}_{\eta}(\epsilon_{\alpha}) + 
X_{\alpha}
\label{eq:deconv}
\end{equation}
for $\alpha=1, \dots, M$, 
where $X_{\alpha}$ represents the unknown errors.
(It should be noted that DDMRG errors $X_{\alpha}$ include significant 
systematic contributions, for instance due to the variational 
nature of the procedure.~\cite{jec02})
In principle, one could determine the true spectrum, i.\,e., the excitation
energies $E^{(N)}_n$ and the corresponding weights
$S^{(N)}_n$, through this system of equations. 
In practice, however, this is an ill-conditioned problem 
except for simple discrete spectra.
Moreover, we are not interested in resolving the discrete peaks of small
systems but in calculating the
piecewise smooth spectra of macroscopic systems.

The spectrum in the thermodynamic limit is given by
\begin{equation}
S(\epsilon) = 
\lim_{\eta \rightarrow 0} \lim_{N \rightarrow \infty} 
S^{(N)}_{\eta}(\epsilon) .
\end{equation}
Note that, generally, the order of the two limits can not be exchanged.
Typically, the spectral function $S(\epsilon)$ is piecewise smooth,
i.\,e., it exhibits one or more continua as well as isolated sharp features 
such as steps, power-law singularities or cusps. 
In principle, one should carry out several DDMRG simulations with varying
system size $N$ and broadening $\eta$ and then extrapolate the numerical
data to obtain $S(\epsilon)$. In most cases, a simultaneous 
extrapolation for $N \rightarrow \infty$ and $\eta \rightarrow 0$ is possible~\cite{jec02}
using a constant value of $\eta N$.
Nevertheless, the computational cost of DDMRG simulations increases very
rapidly with
smaller $\eta$ and the overall cost of this approach is prohibitive for a full spectrum. 
Indeed, this approach  has been mostly used to study isolated spectral features  in the
thermodynamic limit such as power-law singularities and steps.~\cite{jec02,jec08b,jec13}

As all operations used to define $S(\epsilon)$ from $S^{(N)}_{\eta}(\epsilon)$
are linear,
the broadened spectrum of the finite system can also be written explicitly 
as a function
of the infinite system spectrum
\begin{equation}
S^{(N)}_{\eta} (\epsilon) = \int_{-\infty}^{\infty} d\omega \
K^{(N)}_{\eta}(\epsilon,\omega) \ S(\omega) .
\label{eq:spec-inf}
\end{equation}
The kernel $K^{(N)}_{\eta}(\epsilon,\omega)$ includes both the finite-size
effects and the Lorentzian smoothing. Its form is not known but 
it is clear that we must recover a pure Lorentzian smoothing
in the thermodynamic limit
\begin{equation}
\lim_{N\rightarrow \infty} K^{(N)}_{\eta}(\epsilon,\omega)
= L_{\eta}(\epsilon-\omega) .
\label{eq:kernel-limit}
\end{equation}
Combining eqs.~(\ref{eq:deconv}) and~(\ref{eq:spec-inf}) we obtain a system of
equations
\begin{equation}
W^{(N)}_{\eta}(\epsilon_{\alpha})
= \int_{-\infty}^{\infty} d\omega \
K^{(N)}_{\eta}(\epsilon_{\alpha},\omega) \ S(\omega) + X_{\alpha}
\label{eq:inverse-problem}
\end{equation}
for $\alpha = 1, \dots, M$,
relating the DDMRG data set
\begin{equation}
\left \{ \left (\epsilon_{\alpha}, 
W^{(N)}_{\eta}(\epsilon_{\alpha}) \right); \alpha=1,\dots,M \right \}
\label{eq:dmrg-data}
\end{equation}
to the infinite system spectrum $S(\epsilon)$.

Determining $S(\epsilon)$ from these equations is a so-called inverse 
problem.~\cite{nr}
This kind of problem is also called blind deconvolution since
our knowledge of the kernel is incomplete.
[Strictly speaking, it is not a deconvolution because 
eq.~(\ref{eq:inverse-problem}) is not a convolution. However, as
the kernel approaches the form 
$K^{(N)}_{\eta}(\epsilon,\omega) = L_{\eta}(\epsilon-\omega)$
in the thermodynamic limit,
we will use the terminology of deconvolution problems.]
It should be obvious that this is an ill-posed problem. 
First, the errors $X_{\alpha}$ and the kernel
$K^{(N)}_{\eta}(\epsilon,\omega)$ are not known.
Second, the problem is sorely underdetermined as we try 
to reconstruct the function of a continuous variable
from a finite number $M$ of data points. 
Finally, a convolution with a Lorentzian is a smoothing operation and thus the
corresponding deconvolution is an extremely ill-conditioned inverse problem:
the solution will be extremely sensitive to small changes or errors in the
input.

\subsection{Cost function approach}

Various deconvolution methods have been used successfully  to 
deduce piecewise smooth spectra from the broadened finite-system spectra
calculated with DDMRG.
They include, direct inversion at low resolution,~\cite{geb03}
linear regularization methods,~\cite{nis04a,jec07}
Fourier transform with low-pass filtering,~\cite{raa04,raa05}
nonlinear regularization methods such as the
Maximum Entropy Method,~\cite{raa05} 
parametrization with piecewise polynomial functions,~\cite{wei09,raa05}
and a deconvolution ansatz for the self energy.~\cite{sch10,ulb10}
However, this task has not been viewed as a blind deconvolution so far.
Instead, it has been considered as the deconvolution of a perfectly known 
kernel. The need for regularization or filtering techniques
has been viewed as the consequence ill-conditioning and under-determination
of the problem~(\ref{eq:inverse-problem}) with a Lorentzian kernel.

All of these methods offer some advantages for particular spectral forms.
However, their common drawback is that they are ill-suited 
for sharp spectral features, such as steps or power-law singularities,
within or at the edge of a continuum. 
Either the regularization procedure smooths out true sharp features
excessively or it allows the occurrence of deconvolution artifacts
(artificial sharp structures, rapid oscillations or negative spectral 
weight), especially in the vicinity of the true spectrum singularities.
Naturally, better results can be obtained if we can use \textit{a priori}
knowledge about the properties of the spectrum~\cite{whi08,raa05} 
but, in practice, this is a rare occurrence.
Therefore, we need a better method for solving 
the inverse problem~(\ref{eq:inverse-problem}) 
which allows us to determine isolated sharp spectral features
accurately
while preserving the positivity and the piecewise smoothness of $S(\epsilon)$.

Let the DDMRG data (\ref{eq:dmrg-data})
be evenly distributed in the energy interval $[\epsilon_A,\epsilon_B]$.
The difference between two consecutive 
energies is $\Delta\epsilon \sim 1/M$. 
Additionally, consider a set of equidistant energies 
$\{ \omega_{\mu} ; \mu=1,\dots,L \}$
in the interval $[\omega_A,\omega_B] \subset [\epsilon_A,\epsilon_B]$.
The distance between these energies is $\Delta\omega \sim 1/L$. 
As we will always use $L \geq M$, we have $\Delta\omega \leq \Delta\epsilon$.
(Typical values are $M \approx 10^2-10^3$ and $L \approx 10^3-10^4$.)
As in a least-square approach we define a cost function $\chi \left (\{ R_{\mu} \} \right)$
 as the sum of the squares of the differences between the
DDMRG data and an approximate representation parametrized by a discrete set of variables  $\{R_{\mu} ; \mu=1,\dots,L\}$
\begin{eqnarray}
  \chi = \sum_{\alpha=1}^{M} \left ( 
W^{(N)}_{\eta}(\epsilon_{\alpha}) - 
\sum_{\mu=1}^L
K^{(N)}_{\eta}(\epsilon_{\alpha},\omega_{\mu}) R_{\mu} \Delta\omega
\right )^2 .
\label{eq:cost-function}
\end{eqnarray}
The absolute minimum of $\chi \left (\{ R_{\mu} \} \right)$ is zero and
the corresponding parameters $\{R_{\mu}\}$ are determined
by a linear system of $M$ equations  
\begin{equation}
W^{(N)}_{\eta}(\epsilon_{\alpha}) = \sum_{\mu=1}^L 
K^{(N)}_{\eta}(\epsilon_{\alpha},\omega_{\mu}) R_{\mu} \Delta\omega .
\label{eq:linear-system}
\end{equation}
Using this equation system to determine the parameters $\{R_{\mu}\}$ would be an
unconstrained least-square fit.

In the limit $L \rightarrow \infty$ (followed by $\epsilon_{A} 
\rightarrow -\infty$ and $\epsilon_{B} \rightarrow + \infty$)
this equation system becomes equivalent to the inverse
problem~(\ref{eq:inverse-problem}) with vanishing errors $X_{\alpha}=0$.
Thus the absolute minimum of $\chi \left (\{ R_{\mu} \} \right)$ 
yields the spectrum $S(\epsilon)$ through
\begin{equation}
R_{\mu} =  S(\omega_{\mu}) .
\label{eq:solution}
\end{equation}

If we substitute a Lorentz kernel 
$L_{\eta}(\epsilon_{\alpha}-\omega_{\mu})$
for the unknown kernel,
$K^{(N)}_{\eta}(\epsilon_{\alpha},\omega_{\mu})$ 
in~(\ref{eq:linear-system}), 
we recover the finite-system deconvolution problem
defined by equations~(\ref{eq:spec-finite}) and~(\ref{eq:deconv})
for vanishing errors $X_{\alpha}$.
Thus the absolute minimum of the cost function corresponds
to the discrete finite-system spectrum $S^{(N)}(\epsilon)$   
through 
$R_{\mu}\Delta\omega = S^{(N)}_n$ if $\omega_{\mu} = E^{(N)}_n$ and 
$R_{\mu}=0$ otherwise.
Physically, the solution of the deconvolution problem
is unique for vanishing errors $X_{\alpha}$ and thus 
the cost function should have a unique absolute minimum.
From a mathematical point of view, however, the equation
system~(\ref{eq:linear-system}) could have no solution or
infinitely many solutions. Then any small error $X_{\alpha}$ can
generate wildly different (and mostly unphysical) solutions.

Therefore, as~(\ref{eq:kernel-limit}) holds in the thermodynamic limit,
it is possible and preferable to obtain a reasonable approximation of the infinite-system spectrum $S(\epsilon)$
from the minimization of the cost function~(\ref{eq:cost-function}) with a
Lorentz kernel under the constraint that the spectral function
$S(\epsilon)$ is physically allowed. 
For instance, $S(\epsilon)$ should be positive semidefinite and
piecewise smooth.
Generally, this solution does not correspond to the absolute minimum or even a
local minimum of $\chi \left (\{ R_{\mu} \} \right)$. 
Indeed, the solution of the inverse problem~(\ref{eq:inverse-problem})
corresponds to the value
\begin{equation}
\chi \left (\{ R_{\mu} \} \right) = \sum_{\alpha=1}^{M}  X_{\alpha}^2
\end{equation}
if we assume that the relation~(\ref{eq:solution}) holds.
Of course, it could be possible to lower the cost function with other 
configurations
$\{ R_{\mu} \}$ but in that case the relation~(\ref{eq:solution}) would no
longer hold.
Note that, this idea has been implicitly assumed in all previous deconvolution
schemes of DDMRG data aiming at piecewise smooth spectra so far.
However, in these approaches the agreement between solution 
$\{ R_{\mu} \}$ and numerical data $\{ W^{(N)}_{\eta}(\epsilon_{\alpha}) \}$ 
in eq.~(\ref{eq:linear-system}) with a Lorentzian kernel is considered essential
while the regularization of the solution and the errors $X_{\alpha}$
are seen as perturbations
which should deteriorate the agreement as little as possible.

Yet a blind deconvolution requires equal balancing of 
the agreement between solution and numerical data and of 
the smoothness and stability of the solution.~\cite{cam07}
Hence we must take a different point of view: 
The Lorentzian kernel~(\ref{eq:kernel-limit}) is only an approximation
of the true kernel $K^{(N)}_{\eta}(\epsilon,\omega)$ and 
the physical constraints on
the deconvolved spectrum $S(\epsilon)$ are essential in the minimization
of the cost function. 
Thus we accept significant deviations of $\{ R_{\mu} \}$
from the conditions~(\ref{eq:linear-system}) yielding the absolute minimum
of the cost function,
or, equivalently, we assume that the errors $X_{\alpha}$ can be substantial.

\subsection{Method}

Therefore, the blind deconvolution problem can be formulated
as a least-square optimization under non-linear constraints.
We want to minimize the cost function~(\ref{eq:cost-function})
with the Lorentz kernel under the constraints that the spectrum
$S(\epsilon)$ has the following properties:
\begin{enumerate}
\item finite band width, i.e $S(\omega) = 0 $ for all $\omega < \omega_A$ and 
all $\omega > \omega_B$ for some finite $\omega_{A,B}$,
\item positive semi-definite, $S(\epsilon) \geq 0$, and
\item piecewise smooth.
\end{enumerate}
The first two conditions are easily expressed for the parameters
$\{ (\omega_{\mu},R_{\mu}) \}$.
The somewhat fuzzy concept of a piecewise smooth spectrum must now be 
formulated more precisely. In principle, we wish that $S(\epsilon)$ is
piecewise continuous
and that the distance between discontinuities is larger than 
a minimal energy difference $d$. As we must work with a finite number $L$ of 
points
$(\omega_{\mu},R_{\mu})$, 
we only have a discrete representation of $S(\epsilon)$
and we have to formulate a ``continuity'' condition for
the discrete set of variables as well.
Therefore, we require that the distance between 
two significant extrema is larger than a parameter $d > 0$.
Two neighboring extrema at energy $\epsilon_1$ and 
$\epsilon_2$ are significant
if there relative height difference is larger than a parameter $h \geq 0$,
\begin{equation}
2 \frac{\vert S(\epsilon_1)-S(\epsilon_2) \vert}{S(\epsilon_1)+S(\epsilon_2)} 
> h . 
\end{equation}
This condition can easily be formulated for the parameters 
$\{ (\omega_{\mu},R_{\mu}) \}$.
The minimal extremum distance $d$ must be chosen carefully. It should be smaller than
the distance between actual singularities in the spectrum $S(\epsilon)$ 
but a too small value allows many artificial peaks in a deconvolved 
spectrum.
In practice, we have found that we can obtain reasonable solutions
to the blind deconvolution problem which look
piecewise smooth using $d \gtrsim \eta$. In all examples discussed in this paper
every local extremum is considered to be significant (i.\,e., we have used the
precision of floating-point arithmetic $h \approx 10^{-16}$).

The cost function is minimized iteratively.
Iterations are repeated until the procedure converges.
Each iteration consists in two steps. In the first step the cost function
$\chi(\{R_{\mu}\})$ 
is minimized with respect to each variable $R_{\mu} \geq 0$ 
successively.
This minimization under constraint does not present any difficulty as 
$\chi(\{R_{\mu}\})$ is a
second-order polynomial in each variable $R_{\mu}$.
In the second step, we first find the positions
$(\omega_{\nu},\omega_{\tau})$ of 
all significant extrema pairs in $\{(\omega_{\mu},R_{\mu})\}$
which are separated by less than a distance $d$. 
Then we interpolate the data $\{(\omega_{\mu},R_{\mu})\}$ linearly from
$\mu=\nu-1$ to
$\mu=\tau+1$ to smooth out the spectrum around the extrema.
In doing so we take care to preserve the total spectral weight
\begin{equation}
S = \sum_{\mu=1}^{L} R_{\mu} \Delta \omega 
\approx \int_{-\infty}^{+\infty} d\epsilon \ S(\epsilon) .
\end{equation}
The search for extrema and their smoothing is repeated until 
there is no more close significant extrema in $\{ (\omega_{\mu},R_{\mu})\}$. 
Then we start the next iteration.      
By design the first step results in a decrease of $\chi(\{R_{\mu}\})$.
The second step nearly always results in an increase of $\chi(\{R_{\mu}\})$.
Without the second step, however, we would perform an under-determined
($L \gg M$) deconvolution devoid of any regularization mechanism and thus
obtain a completely useless result.
Typically, we observe a rapid and monotonic decrease of 
$\chi(\{R_{\mu}\})$ in the initial iterations followed by a saturation
or oscillations in further iterations. 
Therefore, we monitor the changes in the parameters $R_{\mu}$ and the normalized cost function 
\begin{equation}
\theta^2 = \frac{\chi}{S^2 M}
\end{equation}
to determine converged configurations $\{R_{\mu}\}$.
Convergence requires typically $10^2$ to $10^3$ iterations depending on the
quality of the DDMRG data and the complexity of the spectrum.
Finally, the solution $\{ (\omega_{\mu},R_{\mu}) \}$ can be smoothened
using a narrow Lorentzian distribution to obtain a continuous function
\begin{equation}
S(\epsilon) = \sum_{\mu=1}^{L} R_{\mu} L_{\tilde \eta}(\epsilon -\omega_{\mu})
\label{eq:reconvolution}
\end{equation}
with ${\tilde \eta} \ll \eta$.
Alternatively, we can use a Gaussian distribution~\cite{wei09} of width $\sigma \ll \eta$.
The second approach yields sharper (real or artificial) features because the
tail of a Gaussian distribution decreases faster than that of the Lorentz
distribution.  
As our minimization problem possesses many local minima, the final results $\{R_{\mu}\}$ depend
somewhat on the criteria for convergence. However, if the final smoothening function is broad enough, 
the differences are canceled out.
If the DDMRG data~(\ref{eq:dmrg-data}) 
are not evenly distributed in the interval $[\epsilon_A,\epsilon_B]$ or 
if they already exhibit numerous close extrema, it is useful to 
regularize them
before starting the deconvolution iterations using an interpolation and
the smoothening procedure described above.

The computational effort required by this procedure is negligible compared
to the computational cost of the DDMRG simulations yielding the original data. 
(Our code in the programming language C contains less than 400 lines of
instructions and the deconvolution of one spectrum takes less than 30 minutes
on a single CPU.) 
Therefore, we have not bothered to optimize the algorithm.
Nevertheless, it should be implemented in such a way that it only requires
$\sim L^2$ operations rather than the $\sim L^3$ operations of a straightforward
implementation.

The method described here can be generalized in several ways. For instance, it is possible to use a variable 
spacing $\Delta\epsilon$ of the DDMRG data points, such as a finer mesh close to sharp spectrum features.
However, this does not  seem to improve the results in practice because the broadening parameter $\eta$, not 
$\Delta\epsilon$, is the limiting scale.
A generalization to variable $\eta$ and $\Delta\epsilon$, as proposed in Ref.~\onlinecite{nis04a} for quantum impurity
problems, should also be possible but we have not tested it yet.
To introduce information about the variation of the spectrum with $\eta$ one could combine DDMRG data obtained 
for different values of $\eta$ by defining an overall cost function as the sum of the cost functions for each $\eta$.
These generalizations will be tested in future works.

\section{DOS of one-dimensional Mott insulators}

As an illustration of our deconvolution procedure we discuss its application
to the DOS of one-dimensional paramagnetic Mott insulators. 
The nature of Mott insulators is a long-standing
open problem in the theory of strongly correlated quantum 
systems.~\cite{mot90,geb97}
In a paramagnetic Mott insulator quantum fluctuations or 
frustration of the antiferromagnetic spin exchange coupling
prevents the formation of a long-range magnetic order.
Experimentally, non-magnetic Mott insulators have been found in layered
organic insulators~\cite{kur05} as well as in quasi-one-dimensional cuprate 
chains~\cite{kim04}
and ladders~\cite{azu94}. 
Despite decades of extensive research the properties of Mott insulators,
in particular their single-particle DOS, 
are still poorly understood and thus actively investigated. 
The half-filled Hubbard model~\cite{hub63} 
with repulsive on-site interaction $U\geq 0$ 
is a basic lattice model for 
describing Mott insulators and the Mott metal-insulator transition.
Here we consider the case of the one-dimensional Hubbard model,
which is exactly solvable by Bethe Ansatz.~\cite{lie68, hubbard-book}
At half- filling it describes a paramagnetic Mott insulator with
a charge gap (Mott-Hubbard gap) $2\Delta > 0$ for $U > 0$. 
However, the DOS cannot be calculated directly from the Bethe Ansatz.

All DDMRG spectra used here have been calculated with a variable number of
density-matrix eigenstates kept (up to 512) to reach a discarded weight
lower than $10^{-4}$ and to check DMRG truncation errors.
Typically, convergence was reached after three sweeps for each frequency interval of size $\eta$. The DDMRG method is
presented in detail in Ref.~\onlinecite{jec02}.

For $U=0$  the exact DOS of the tight-binding chain  in the thermodynamic limit is
\begin{equation}
D_{\text{tb}}(\epsilon) = \frac{1}{\pi} \frac{1}{\sqrt{4t^2 - \epsilon^2}}
\label{eq:dos-tb}
\end{equation}
for $|\epsilon| < 2t$ while $D_{\text{tb}}(\epsilon)$ vanishes for larger $\vert \epsilon \vert $.
For finite coupling $U$ the spectrum consists in two symmetric Hubbard bands
separated by the gap $2\Delta$.
Low-order strong-coupling perturbation theory~\cite{scpt1,scpt2}   predicts 
a square-root divergence at the DOS threshold for $U \gg 4t$, namely
\begin{equation}
D_{\text{scpt}}(\epsilon) = \frac{1}{2} D_{\text{tb}} \left ( \vert \epsilon \vert -\frac{U}{2} 
\right )
\label{eq:dos-scpt}
\end{equation}
for $\left \vert \vert \epsilon \vert -\frac{U}{2} \right \vert < 2t$ 
and $D_{\text{scpt}}(\epsilon) = 0$
otherwise with $\Delta = \frac{U}{2}-2t$.
Because of this strong-coupling result and the result of the Hartree-Fock (HF) approximation
it has often been assumed that the DOS of one-dimensional Mott insulators
exhibits a square-root divergence at the spectrum onset $\epsilon=\pm\Delta$ like in a one-dimensional band insulator.
Indeed, in the unrestricted HF approximation
the one-dimensional half-filled Hubbard model is an antiferromagnetic
Mott insulator for $U > 0$. Its DOS is given by 
\begin{equation}
D(\epsilon) = \frac{1}{\pi} 
\frac{\epsilon}{\sqrt{\left ( \epsilon^2 - \Delta_{\text{HF}}^2 \right ) 
\left ( 4t^2 + \Delta_{\text{HF}}^2 - \epsilon^2 \right ) }} 
\end{equation}
for $\Delta_{\text{HF}} < |\epsilon| < \sqrt{4t^2+\Delta_{\text{HF}}^2}$
and vanishes otherwise.
Here $2\Delta_{\text{HF}}$ is the HF gap. For $U=0$, $\Delta_{\text{HF}}=0$ and this DOS reduces
to the DOS of the tight-biding chain~(\ref{eq:dos-tb}).
For $U>0$, $\Delta_{\text{HF}} \neq 0$ and the HF DOS shows a square-root divergence at 
the onset of the spectrum.
However, in the weak-coupling limit $U \ll 4t$ 
a field-theoretical analysis~\cite{ess02} predicts 
that the DOS of one-dimensional Mott insulators is constant above the threshold energy $\Delta$,
\begin{equation}
D_{\text{ft}}(\epsilon) = C \ \theta(|\epsilon| - \Delta)
\label{eq:dos-ft}
\end{equation}
at least for $\vert \epsilon\vert \leq 3 \Delta \ll 4t$.
Thus there is a discrepancy between the field-theoretical 
and strong-coupling predictions for the behavior of $D(\epsilon)$ just above 
the threshold energy $\Delta$.

\begin{figure}
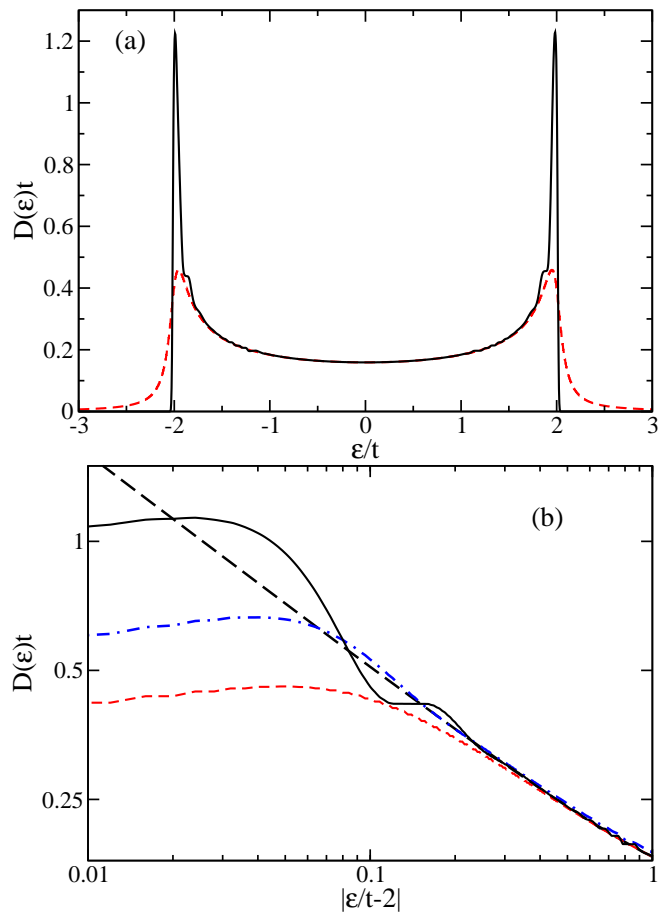

\includegraphics[width=0.48\textwidth]{fig1a.eps}
\includegraphics[width=0.48\textwidth]{fig1b.eps}
\caption{\label{fig:tb} (Color online) DOS of a tight-binding chain:
(a) DDMRG spectrum
for a $128$-site chain with a Lorentzian broadening $\eta=0.08t$ 
(red dashed line) and 
result of our deconvolution method with a Gaussian broadening 
$\sigma = \eta/10$ (black line).
(b) Enlarged view close to the singularity at $\epsilon=2t$ on a double
logarithmic scale: Exact results (black long-dashed line), DDMRG data (red
short-dashed line),
and the deconvolved spectra $\{ (\omega_{\mu},R_{\mu})\}$ for 
$\theta \approx 6\cdot10^{-4}$ (black solid line) and $\theta \approx 8\cdot10^{-3}$
(blue dash-dot line).}
\end{figure}

Figure~\ref{fig:tb}(a) shows the DOS of a tight-binding chain calculated with DDMRG
and the result of our deconvolution procedure.
The DDMRG spectrum has been calculated
in the middle of an open chain with $N=128$ sites using a broadening $\eta=0.08t$, which is
just broad enough to hide its discreteness.
We see that the square-root divergences at $\epsilon=\pm 2t$ have been
smoothed into two broad peaks and that there is substantial spectral weight
at energies $\vert \epsilon \vert  > 2t$.
The deconvolved DOS has been determined from these same DDMRG data
using a minimal extremum distance $d = 2\eta=0.16t$ and a final Gaussian broadening with $\sigma = \eta/10 = 0.008t$.
We see now that the singularities at $\epsilon=\pm 2t$ are clearly visible as sharp peaks and that there is not any
spectral weight at $\vert \epsilon \vert  > 2t$.
Overall the deconvolved DOS is in excellent agreement with the exact spectrum
in the thermodynamic limit~(\ref{eq:dos-tb}).
In particular, we do not observe  any unphysical artefact
such as negative spectral weight.

However, in fig.~\ref{fig:tb}(a) we observe two shoulders in the deconvolved DOS at energies $\epsilon \approx \pm 1.8t$, 
which are not present in the exact solution~(\ref{eq:dos-tb}). An enlarged view close to the singularity at 
$\epsilon = +2t$ is shown in fig.~\ref{fig:tb}(b) on a double logarithmic 
scale. 
We see that the DDMRG data agree with the exact result  only at some distance from the singularity.
In this figure we also show deconvolved spectra $\{ (\omega_{\mu},R_{\mu})\}$ for two different
values of the normalized cost function $\theta$. Clearly, they reproduce the square-root divergence
at $\epsilon = +2t$ much better than the original DDMRG data. The overall divergent behavior is visible
on a broader energy scale for the smaller value of $\theta$ but we see that the reduction of the cost function
is also accompanied by stronger oscillations around the exact result. These oscillations correspond
to the shoulder seen in fig.~\ref{fig:tb}(a).

The occurrence of artificial shoulder-like structures is the main drawback of
our deconvolution procedure.
Any deconvolution method magnifies the noise (numerical errors) which is
present in the original data. This the main issue that existing
methods try to solve in different 
ways.~\cite{wei09,geb03,nis04a,nis04b,raa04,raa05,jec07,sch10,ulb10} 
We have systematically tested our deconvolution procedure using exact results
for non-interacting systems and purposely adding random numerical errors.
We have found that 
by preventing the formation of local maxima in the deconvolved spectrum
our procedure allows us to control the noise magnification 
only partially.
Unfortunately, it cannot handle extrema ($\equiv$ oscillations) in 
the spectrum derivative.
Thus the magnified noise shows up as shoulder-like
structures (but not as local maxima, discontinuities, or sharp angles) 
on an energy scale $d$ and gives a rough appearance to some
deconvoled spectra presented here.
In principle, we should be able to correct this deficiency with a higher-order
interpolation procedure in the smoothening step or 
with a smoothening of the derivative of $S(\epsilon)$
(i.\,e., the finite differences between the parameters $R_{\mu}$). 
However, we have not yet succeeded
in developing a practical algorithm based on these ideas.

\begin{figure}
\includegraphics[width=0.48\textwidth]{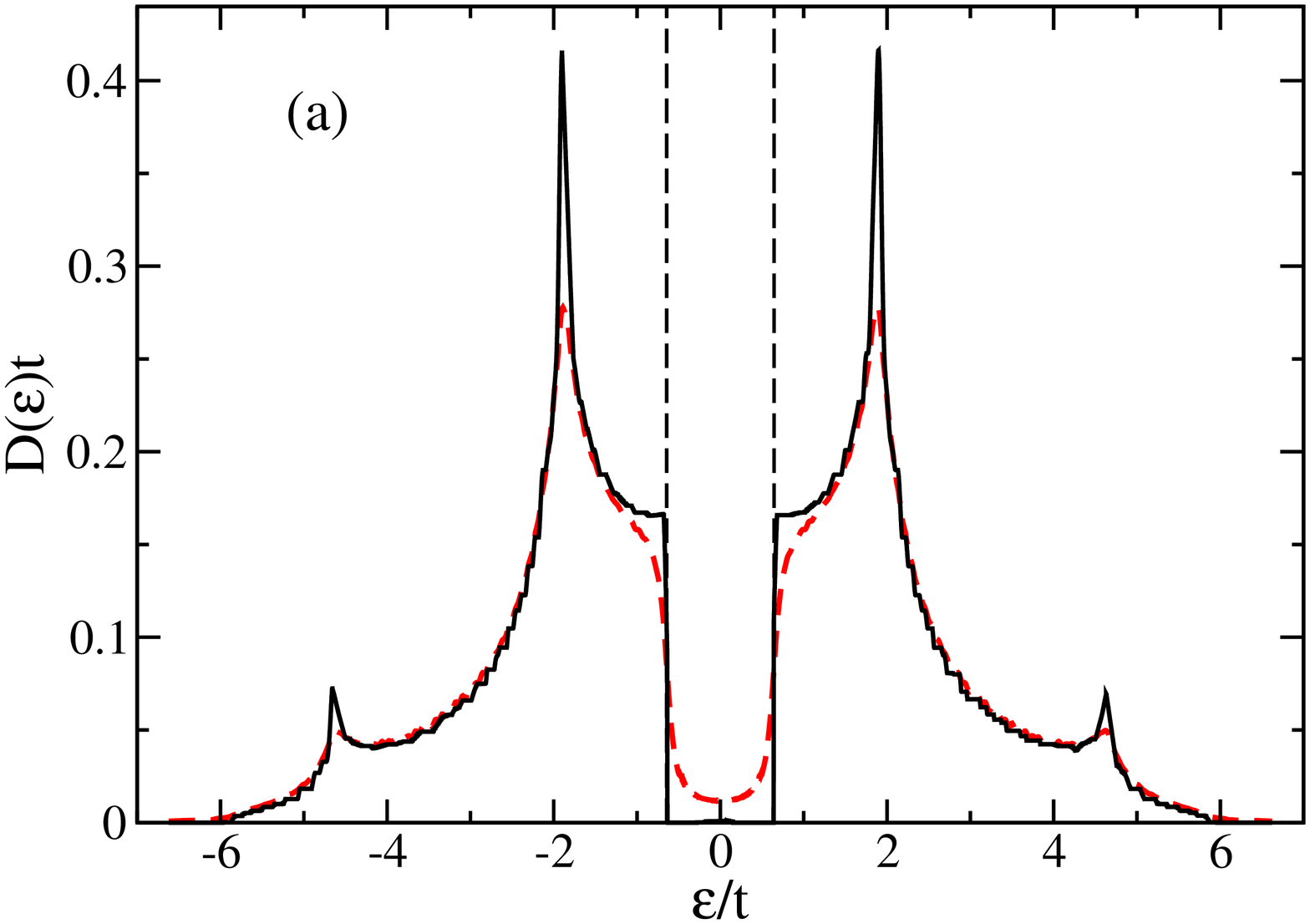}
\includegraphics[width=0.48\textwidth]{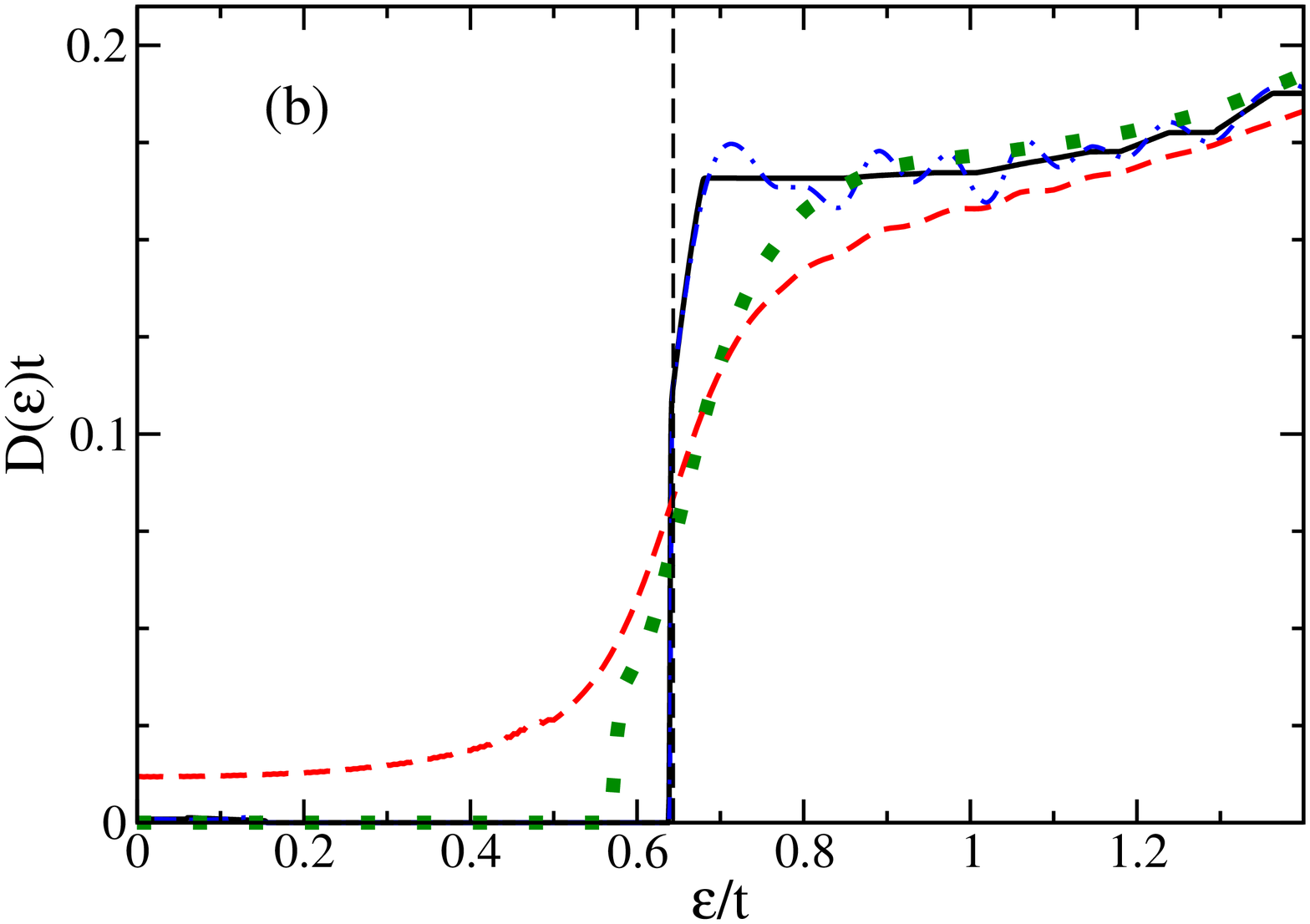}
\caption{\label{fig:onedim4} (Color online)
(a) DOS of the one-dimensional half-filled Hubbard 
model at $U=4t$ 
calculated with DDMRG in a 128-site chain using a broadening $\eta=0.08t$
(red dashed line) and the result of our deconvolution procedure
(black line) with a minimal extremum distance $d=2\eta$ and a Gaussian broadening $\sigma=\eta/100$. 
The vertical dashed lines show the exact position of the
DOS threshold calculated from the Bethe Ansatz solution.
(b) Enlarged view of the same data around the DOS threshold for $\epsilon >0$.
Additionally, the result of our procedure with $d=\eta/2$ (blue dot-dash
line) and of a deconvolution with linear 
regularization~\cite{nr,nis04a,jec07} (green dots) are also shown.}
\end{figure}

Figure~\ref{fig:onedim4}(a) shows the single-particle DOS of the half-filled
one-dimensional Hubbard model at $U=4t$  calculated with DDMRG and the result
of our deconvolution procedure. 
The DDMRG spectrum has been computed~\cite{jec08b} in the middle of an open chain with 
$N=128$ sites and a broadening $\eta=0.08t$.
The deconvolved spectrum has been obtained from these DDMRG data using 
a minimal extremum distance $d=2\eta=0.16t$
and a final Gaussian broadening $\sigma = \eta/100=8\cdot10^{-4}t$.
The effects of the broadening are clearly visible in the DDMRG data. 
For instance, although one can recognize the opening of the Mott-Hubbard gap $2\Delta$,
spectral weight is clearly visible inside the gap. 
A point-wise analysis~\cite{jec08b} of the scaling for $N \rightarrow \infty$ 
with $\eta N=10.24t$ is required to confirm that
the spectral weight jumps from $0$ to a finite value at the onset
$\epsilon = \Delta$ and that the gap width
agrees with the exact results $E_c=2\Delta = 1.286t$ 
calculated with the Bethe Ansatz.~\cite{lie68,hubbard-book}
However, the behavior for $\epsilon \gtrsim \Delta$ remains uncertain because
of the relatively large broadening used in the DDMRG calculation.
On the contrary the deconvolved DOS clearly shows a gap with the
step-like onset~(\ref{eq:dos-ft}) predicted by field theory~\cite{ess02}
at the position $\epsilon=\Delta$ given by the Bethe Ansatz solution.
We obtain similarly unambiguous results for $U$ up to $8t$ (see below).
Therefore, our numerical investigation confirms the field-theoretical prediction~\cite{ess02}
for the onset of the DOS in one-dimensional Mott insulators. 

The superiority of the deconvolved spectrum over the original DDMRG data is
even more obvious in fig.~\ref{fig:onedim4}(b) which shows an enlarged view of the 
DOS around $\epsilon=\Delta$. 
In this figure we also show the result of a deconvolution of the DDMRG data with
a standard linear regularization method~\cite{nr,nis04a,jec07}.
(Note that this method yields negative spectral weights for some energies $\epsilon$
but we show the positive parts only.) We see that the result of the deconvolution procedure
proposed in this work is much superior to that of the standard one, which is too blurred
to allow us to determine the true form of the DOS at the onset $\epsilon=\Delta$.
In addition, fig.~\ref{fig:onedim4}(b) shows the result of our deconvolution procedure 
for a minimal extremum distance $d=\eta/2=0.04t$ which is deliberately too small. In that case, 
artificial oscillations on energy scales $\epsilon \alt \eta$ are clearly visible
in the deconvolved spectrum.

\begin{figure}
\includegraphics[width=0.48\textwidth]{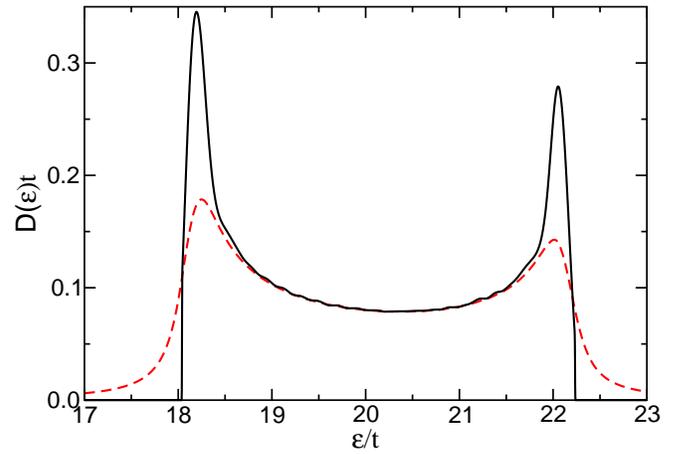}
\caption{\label{fig:onedim40} (Color online)
Upper Hubbard band in the
DOS of the one-dimensional half-filled Hubbard model at $U=40t$ 
calculated with DDMRG in a 64-site chain using a broadening $\eta=0.16t$
(red dashed line) and the result of our deconvolution procedure
(black line) with a Gaussian broadening $\sigma = \eta/100$. 
}
\end{figure}

In the strong-coupling limit  $U \gg t$ our results are less conclusive.
For instance,  we show the DDMRG and deconvolved upper Hubbard and
 for a very strong coupling $U=40t$ in fig.~\ref{fig:onedim40}.
The DDMRG spectrum has been calculated in the middle of an open chain with 
$N=64$ sites using a broadening $\eta=0.16t$.
The deconvolved spectrum has been obtained from these DDMRG data using
a minimal extremum distance $d=\eta$ and a final 
a Gaussian broadening $\sigma = \eta/100$.
Two broad peaks are clearly visible at energies $\epsilon \approx \frac{U}{2} \pm 2t$ as predicted for the
strong-coupling limit. However, the widths and heights of these peaks after deconvolution
are not compatible with the square-root divergences~(\ref{eq:dos-scpt}) predicted by the low-order strong-coupling expansion.
Actually, higher-order corrections indicate~\cite{scpt1,penc} that some spectral weight is present below the peak
at $\epsilon \approx \frac{U}{2} - 2t = 18t$ 
on a scale set by the effective spin exchange coupling
$J=4t^2/U = 0.1t$. This width is compatible with our deconvolved spectra for $U\geq 16t$.
Thus our numerical results agree at least qualitatively with strong-coupling perturbation
theory and suggest
that the square-root divergences in the DOS~(\ref{eq:dos-scpt}) is an artifact of a truncated
strong-coupling expansion.

Nevertheless, 
for strong coupling such as $U=40t$ we are not able to determine the shape of the DOS at the onset
$\epsilon=\Delta$. In particular, it is not clear if the field-theoretical prediction~(\ref{eq:dos-ft}) is still
valid.
Indeed, if the distance $\sim J=4t^2/U$ between onset at $\epsilon = \Delta$ and 
peak at $\epsilon \approx \frac{U}{2} - 2t$ becomes smaller than the minimal extremum
distance $d$, we can no longer distinguish both structures in the deconvolved spectrum.
 In practice, the distance
$d$ must be comparable to the broadening $\eta$ of the DDMRG data
to obtain piecewise smooth spectra. 
Therefore,  although we can
reduce the broadening of isolated spectral structures by several orders of magnitude,
we cannot resolve distinct spectral features on a scale lower than the original broadening of the DDMRG data.
This is a limitation of our deconvolution method that one has to keep in mind.

\begin{figure}
\includegraphics[width=0.48\textwidth]{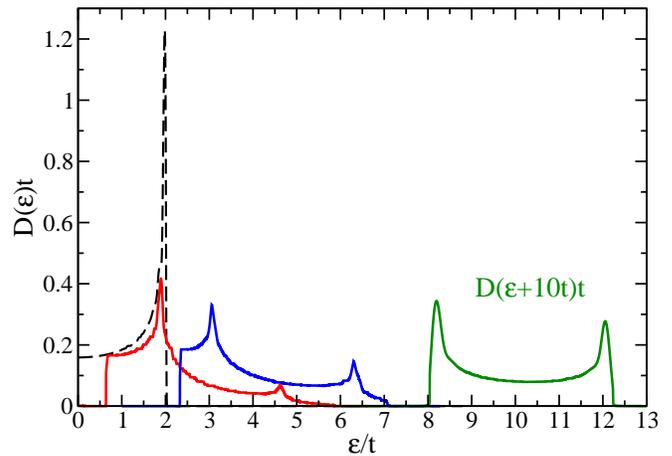}
\caption{\label{fig:onedim} (Color online)
Deconvolved DOS of the half-filled Hubbard model 
for energies $\epsilon > 0$ 
for $U=0$ (dashed line), and from left to right $U/t=4$ (red line), $8$ (blue line) and $40$ (green line). 
The result for $U=40t$ is shifted
to the left by $10t$. } 
\end{figure}

Finally,  fig.~\ref{fig:onedim} recapitulates the evolution of the DOS as a function of the interaction strength
$U/t$. As the spectrum is symmetric $D(-\epsilon) = D(\epsilon)$, we show only the deconvolved spectra 
for positive energies
(i.\,e., the upper Hubbard band). 
At $U=0$ the spectrum consists in a single band with two clearly visible square-root singularities at the band edges
$\epsilon = \pm 2t$. For weak coupling $U$ the band splits into two symmetric Hubbard bands separated
by a gap $2\Delta$, which agrees perfectly with the charge gap calculated from the Bethe Ansatz solution.
At the DOS onsets $\epsilon=\pm \Delta$ the spectrum exhibits the step-like behavior~(\ref{eq:dos-ft})
predicted by field theory.~\cite{ess02} There is an apparent plateau between the onset and a strong first peak,
which evolves from the square-root singularities at  $\epsilon = \pm 2t$ for $U=0$.  Additionally, we observe substantial spectral
weight and a small second peak at higher excitation energy. However, for weak enough $U$ most of the spectral weight
lies between the spectrum onset and the first peak.

As $U$ increases (compare the spectra for $U=4t$ and $U=8t$ in fig.~\ref{fig:onedim}), 
the spectrum and all its features shift to higher excitation energy and the spectral weight becomes more concentrated between
 the visible peaks. In addition, we note
that the separation between onset energy $\Delta$ and the strong first peak becomes systematically smaller until
it is no longer resolvable with our method,  the peak separation  increases
monotonically from about $2t$ for $U\ll t$ to approximately $4t$ for $U\gg t$, and the strength of both peaks become
more equal. 

Comparing the DOS with the momentum-resolved spectral function and the Bethe Ansatz dispersion
(see figs. 4 and 5 in Ref.~\onlinecite{jec08b}) we note that the strong first peak corresponds to the edge
of the spinon branch at momentum $k=0$, the weak second peak corresponds to the edge of the holon branch
at $k=\pm \frac{\pi}{2}$, and the upper edge of the DOS spectrum coincide with the edge of the single spinon-holon
continuum.
Finally, we do not observe any spectral weight outside the first lower and upper Hubbard bands and these two bands 
account for the full spectral weight~(\ref{eq:sum}). Thus we conclude that higher-energy Hubbard bands do not carry
any spectral weight in the bulk single-particle DOS.

\section{Conclusion}

We have presented a blind deconvolution procedure which allows us
to obtain piecewise smooth spectral functions for infinite-size systems
from the DDMRG spectra of finite systems.
It involves a trade-off between the agreement of the deconvolved spectrum
to the original DDMRG data and the piecewise smoothness and positivity of spectral functions.
In practice, the method reduces to a least-square optimization  
under non-linear constraints which enforce the positivity and
piecewise smoothness.
We have tested this deconvolution method on many spectra which are known exactly
in the thermodynamic limit, such as the single-particle density of states and the optical conductivity
of correlated one-dimensional insulators.~\cite{jec00,jec02,ess01}
We have found that our method works well for several kinds of singularities
(e.\,g.\ power-law band edges, steps, excitonic peaks) 
in piecewise smooth spectra.
In particular, it allows us to reduce the broadening by orders of magnitude
and even to substitute the Lorentzian broadening by a Gaussian one.
Its main drawback is the frequent appearance of artificial shoulder-like structures
on energy scales $\Delta \epsilon \gtrsim d$.

We have demonstrated the deconvolution procedure on the single-particle DOS in the one-dimensional
Hubbard model at half filling. Our results show that the DOS has a step-like
shape but no square-root singularity at the spectrum onset in agreement with a field-theoretical prediction for one-dimensional 
paramagnetic Mott
insulators.~\cite{ess02}  In addition, the deconvolution procedure has allowed us to detail the evolution of the
DOS from the non-interacting limit $U=0$ to the strong-coupling limit $U \gg t$.


\begin{acknowledgments}
We thank Karlo Penc and Fabian E\ss ler  for helpful discussions.
The GotoBLAS library developed by Kazushige Goto was used to perform
the DDMRG calculations. Some of these calculations were
carried out on the RRZN cluster system of the Leibniz
Universit\"{a}t Hannover.
\end{acknowledgments}


\end{document}